\documentstyle[12pt]{article}
\def\be {\begin{equation}}
\def\ee {\end{equation}}
\def\ba {\begin{eqnarray}}
\def\ea {\end{eqnarray}}

\def\bi {\begin{itemize}}
\def\ei {\end{itemize}}
\begin{document}
\def\bea{\begin{eqnarray}}
\def\eea{\end{eqnarray}}
\title{\bf  {Interacting  holographic dark energy model in non-flat universe}}
 \author{M.R. Setare  \footnote{E-mail: rezakord@ipm.ir}
  \\{Department of Science,  Payame Noor University. Bijar. Iran}}
\date{\small{}}

\maketitle
\begin{abstract}
We employ
 the holographic model of interacting dark energy to obtain the equation of state  for the holographic energy density
 in non-flat (closed) universe enclosed by
 the event horizon measured from the
 sphere of horizon named $L$.
 \end{abstract}

\newpage

\section{Introduction}
The accelerated expansion that based on recent astrophysical data
\cite{exp}, our universe is experiencing  is today's most
important problem of cosmology. Missing energy density - with
negative pressure - responsible for this expansion has been dubbed
Dark Energy (DE). Wide range of scenarios have been proposed to
explain this acceleration while most of them can not explain all
the features of universe or they have so many parameters that
makes them difficult to fit. The models which have been discussed
widely in literature are those which consider vacuum energy
(cosmological constant) \cite{cosmo} as DE, introduce fifth
elements and dub it quintessence \cite{quint} or scenarios named
phantom \cite{phant} with $w<-1$ , where $w$ is parameter of
state.

An approach to the problem of DE arises from holographic principle
that states that the number of degrees of freedom related directly
to entropy scales with the enclosing area of the system. It was
shown by 'tHooft and Susskind \cite{hologram} that effective local
quantum field theories greatly overcount degrees of freedom because
the entropy scales extensively for an effective quantum field theory
in a box of size $L$ with UV cut-off $ \Lambda$. As pointed out by
\cite{myung}, attempting to solve this problem, Cohen {\it et al.}
showed \cite{cohen} that in quantum field theory, short distance
cut-off $\Lambda$ is related to long distance cut-off $L$ due to the
limit set by forming a black hole. In other words the total energy
of the system with size $L$ should not exceed the mass of the same
size black hole i.e. $L^3 \rho_{\Lambda}\leq LM_p^2$ where
$\rho_{\Lambda}$ is the quantum zero-point energy density caused by
UV cutoff $\Lambda$ and $M_P$ denotes Planck mass ( $M_p^2=1/{G})$.
The largest $L$ is required to saturate this inequality. Then its
holographic energy density is given by $\rho_{\Lambda}= 3c^2M_p^2/
L^2$ in which $c$ is free dimensionless parameter and coefficient 3
is for convenience.

 As an application of holographic principle in cosmology,
 it was studied by \cite{KSM} that consequence of excluding those degrees of freedom of the system
 which will never be observed by that effective field
 theory gives rise to IR cut-off $L$ at the
 future event horizon. Thus in a universe dominated by DE, the
 future event horizon will tend to constant of the order $H^{-1}_0$, i.e. the present
 Hubble radius. The consequences of such a cut-off could be
 visible at the largest observable scales and particulary in the
 low CMB multipoles where we deal with discrete wave numbers. Considering the power spectrum in finite
 universe as a consequence of holographic constraint, with different boundary
 conditions, and fitting it with LSS, CMB and supernova data, a cosmic duality between dark energy equation of state
 and power spectrum is obtained that can describe the low $l$ features extremely
 well.

 Based on cosmological state of holographic principle, proposed by Fischler and
Susskind \cite{fischler}, the Holographic model of Dark Energy (HDE)
has been proposed and studied widely in the
 literature \cite{miao,HDE}. In \cite{HG} using the type Ia
 supernova data, the model of HDE is constrained once
 when c is unity and another time when c is taken as free
 parameter. It is concluded that the HDE is consistent with recent observations, but future observations are needed to
 constrain this model more precisely. In another paper \cite{HL},
 the anthropic principle for HDE is discussed. It is found that,
 provided that the amplitude of fluctuation are variable the
 anthropic consideration favors the HDE over the cosmological
 constant.

 In HDE, in order to determine the proper and well-behaved system's IR cut-off, there are some
difficulties that must be studied carefully to get results adapted
with experiments that claim our universe has accelerated
expansion. For instance, in the model proposed by \cite{miao}, it
is discussed that considering particle horizon, $R_p$,
 \be
 R_p=a\int_0^t\frac{dt}{a}=a\int_0^a\frac{da}{Ha^2}
\ee
 as the IR cut-off, the HDE density reads to be
 \be
  \rho_{\Lambda}\propto a^{-2(1+\frac{1}{c})},
\ee
 that implies $w>-1/3$ which does not lead to accelerated
universe. Also it is shown in \cite{easther} that for the case of
closed
universe, it violates the holographic bound.\\

The problem of taking apparent horizon (Hubble horizon) - the
outermost surface defined by the null rays which instantaneously
are not expanding, $R_A=1/H$ - as the IR cut-off in the flat
universe, was discussed by Hsu \cite{Hsu}. According to Hsu's
argument, employing Friedman equation $\rho=3M^2_PH^2$ where
$\rho$ is the total energy density and taking $L=H^{-1}$ we will
find $\rho_m=3(1-c^2)M^2_PH^2$. Thus either $\rho_m$ and
$\rho_{\Lambda}$ behave as $H^2$. So the DE results pressureless,
since $\rho_{\Lambda}$ scales as like as matter energy density
$\rho_m$ with the scale factor $a$ as $a^{-3}$. Also, taking
apparent horizon as the IR cut-off may result the constant
parameter of state $w$, which is in contradiction with recent
observations implying variable $w$ \cite{varw}. In our
consideration for non-flat universe, because of the small value of
$\Omega_k$ we can consider our model as a system which departs
slightly from flat space. Consequently we respect the results of
flat universe so that we treat apparent horizon only as an
arbitrary distance and not as the system's IR cut-off.

 On the other hand taking the event horizon, $R_h$, where
 \be
  R_h= a\int_t^\infty \frac{dt}{a}=a\int_a^\infty\frac{da}{Ha^2}
 \ee
 to be the IR cut-off, gives the results compatible with observations for flat
 universe.

 It is fair to claim that simplicity and reasonability of HDE provides
 more reliable frame to investigate the problem of DE rather than other models
proposed in the literature\cite{cosmo,quint,phant}. For instance the
coincidence or "why now" problem is easily solved in some models of
HDE based on this fundamental assumption that matter and holographic
dark energy do not conserve separately, but the matter energy
density decays into the holographic energy density \cite{interac}.
In fact a suitable evolution of the Universe is obtained when, in
addition to the holographic dark energy, an interaction (decay of
dark energy to matter) is assumed.\\
Some experimental data has implied that our universe is not a
perfectly flat universe and recent papers have favored the universe
with spatial curvature \cite{{wmap},{ws}}. As a matter of fact, we
want to remark that although it is believed that our universe is
flat, a contribution to the Friedmann equation from spatial
curvature is still possible if the number of e-foldings is not very
large \cite{miao2}. Defining the appropriate distance, for the case
of non-flat universe has another story. Some aspects of the problem
has been discussed in \cite{miao2,guberina}. In this case, the event
horizon can not be considered as the system's IR cut-off, because
for instance, when the dark energy is dominated and $c=1$, where $c$
is a positive constant, $\Omega_\Lambda=1+ \Omega_k$, we find $\dot
R_h<0$, while we know that in this situation we must be in de Sitter
space with constant EoS. To solve this problem, another distance is
considered- radial size of the event horizon measured on the sphere
of the horizon, denoted by $L$- and the evolution of holographic
model of dark energy in non-flat universe is investigated.
\\
In present paper, using the holographic model of dark energy in
non-flat universe, we obtain equation of state for interacting
holographic dark energy density in a universe enveloped by  $L$ as
the system's IR cut-off.
\section{ Intracting holographic dark energy density }
In this section we obtain the equation of state for the holographic
energy density when there is an interaction between holographic
energy density $\rho_{\Lambda}$ and a Cold Dark Matter(CDM) with
$w_{m}=0$. The continuity equations for dark energy and CDM are
\begin{eqnarray}
\label{2eq1}&& \dot{\rho}_{\rm \Lambda}+3H(1+w_{\rm \Lambda})\rho_{\rm \Lambda} =-Q, \\
\label{2eq2}&& \dot{\rho}_{\rm m}+3H\rho_{\rm m}=Q.
\end{eqnarray}
The interaction is given by the quantity $Q=\Gamma
\rho_{\Lambda}$. This is a decaying of the holographic energy
component into CDM with the decay rate $\Gamma$. Taking a ratio
of two energy densities as $r=\rho_{\rm m}/\rho_{\rm \Lambda}$,
the above equations lead to
\begin{equation}
\label{2eq3} \dot{r}=3Hr\Big[w_{\rm \Lambda}+
\frac{1+r}{r}\frac{\Gamma}{3H}\Big]
\end{equation}
 Following Ref.\cite{Kim:2005at},
if we define
\begin{eqnarray}\label{eff}
w_\Lambda ^{\rm eff}=w_\Lambda+{{\Gamma}\over {3H}}\;, \qquad w_m
^{\rm eff}=-{1\over r}{{\Gamma}\over {3H}}\;.
\end{eqnarray}
Then, the continuity equations can be written in their standard
form
\begin{equation}
\dot{\rho}_\Lambda + 3H(1+w_\Lambda^{\rm eff})\rho_\Lambda =
0\;,\label{definew1}
\end{equation}
\begin{equation}
\dot{\rho}_m + 3H(1+w_m^{\rm eff})\rho_m = 0\; \label{definew2}
\end{equation}
We consider the non-flat Friedmann-Robertson-Walker universe with
line element
 \be\label{metr}
ds^{2}=-dt^{2}+a^{2}(t)(\frac{dr^2}{1-kr^2}+r^2d\Omega^{2}).
 \ee
where $k$ denotes the curvature of space k=0,1,-1 for flat, closed
and open universe respectively. A closed universe with a small
positive curvature ($\Omega_k\sim 0.01$) is compatible with
observations \cite{ {wmap}, {ws}}. We use the Friedmann equation to
relate the curvature of the universe to the energy density. The
first Friedmann equation is given by
\begin{equation}
\label{2eq7} H^2+\frac{kc^2}{a^2}=\frac{1}{3M^2_p}\Big[
 \rho_{\rm \Lambda}+\rho_{\rm m}\Big].
\end{equation}
Define as usual
\begin{equation} \label{2eq9} \Omega_{\rm
m}=\frac{\rho_{m}}{\rho_{cr}}=\frac{ \rho_{\rm
m}}{3M_p^2H^2},\hspace{1cm}\Omega_{\rm
\Lambda}=\frac{\rho_{\Lambda}}{\rho_{cr}}=\frac{ \rho_{\rm
\Lambda}}{3M^2_pH^2},\hspace{1cm}\Omega_{k}=\frac{kc^2}{a^2H^2}
\end{equation}
Now we can rewrite the first Friedmann equation as
\begin{equation} \label{2eq10} \Omega_{\rm m}+\Omega_{\rm
\Lambda}=1+\Omega_{k}.
\end{equation}
Using Eqs.(\ref{2eq9},\ref{2eq10}) we obtain following relation
for ratio of energy densities $r$ as
\begin{equation}\label{ratio}
r=\frac{1+\Omega_{k}-\Omega_{\Lambda}}{\Omega_{\Lambda}}
\end{equation}
In non-flat universe, our choice for holographic dark energy
density is
 \be \label{holoda}
  \rho_\Lambda=3c^2M_{p}^{2}L^{-2}.
 \ee
As it was mentioned, $c$ is a positive constant in holographic
model of dark energy($c\geq1$)and the coefficient 3 is for
convenient. $L$ is defined as the following form:
\begin{equation}\label{leq}
 L=ar(t),
\end{equation}
here, $a$, is scale factor and $r(t)$ can be obtained from the
following equation
\begin{equation}\label{rdef}
\int_0^{r(t)}\frac{dr}{\sqrt{1-kr^2}}=\int_t^\infty
\frac{dt}{a}=\frac{R_h}{a},
\end{equation}
where $R_h$ is event horizon. Therefore while $R_h$ is the radial
size of the event horizon measured in the $r$ direction, $L$ is the
radius of the event horizon measured on the sphere of the horizon.
\footnote{ As I have discussed in introduction, in non-flat case the
event horizon can not be considered as the system's IR cut-off,
because if we use $R_h$ as IR cut-off, the holographic dark energy
density is given by \be
  \rho_\Lambda=3c^2M_{p}^{2}R_{h}^{-2}.
 \ee
 When there is only dark energy and the curvature,
$\Omega_\Lambda=1+ \Omega_k$, and $c=1$, we find
 \cite{{miao2}} \be \dot R_h=\frac{1}{\sqrt{\Omega_{\Lambda}}}-1=\frac{1}{\sqrt{1+\Omega_{k}}}-1<0, \ee while we know
that in this situation we must be in de Sitter space with constant
EoS.} For closed universe we have (same calculation is valid for
open universe by transformation)
 \be \label{req}
 r(t)=\frac{1}{\sqrt{k}} sin y.
 \ee
where $y\equiv \sqrt{k}R_h/a$. Using definitions
$\Omega_{\Lambda}=\frac{\rho_{\Lambda}}{\rho_{cr}}$ and
$\rho_{cr}=3M_{p}^{2}H^2$, we get

\begin{equation}\label{hl}
HL=\frac{c}{\sqrt{\Omega_{\Lambda}}}
\end{equation}
Now using Eqs.(\ref{leq}, \ref{rdef}, \ref{req}, \ref{hl}), we
obtain \footnote{Now we see that the above problem is solved when
$R_h$ is replaced with $L$. According to eqs.(\ref{2eq9},
\ref{holoda}), the ratio of the energy density between curvature and
holographic dark energy is \be \label{ratio}
\frac{\Omega_{k}}{\Omega_{\Lambda}}=\frac{\sin^{2}y}{c^2}
 \ee
when there is only dark energy and the curvature, $\Omega_\Lambda=1+
\Omega_k$, and $c=1$, we find $\Omega_\Lambda=\frac{1}{\cos^{2}y}$,
in this case according to eq.(\ref{ldot}) $\dot L=0$, therefore, as
one expected in this de Sitter space case, the dark energy remains a
constant.
 }
 \be \label{ldot}
 \dot L= HL+ a \dot{r}(t)=\frac{c}{\sqrt{\Omega_\Lambda}}-cos y,
\end{equation}
By considering  the definition of holographic energy density
$\rho_{\rm \Lambda}$, and using Eqs.( \ref{hl}, \ref{ldot}) one
can find:
\begin{equation}\label{roeq}
\dot{\rho_{\Lambda}}=-2H(1-\frac{\sqrt{\Omega_\Lambda}}{c}\cos
y)\rho_{\Lambda}
\end{equation}
Substitute this relation into Eq.(\ref{2eq1}) and using
definition $Q=\Gamma \rho_{\Lambda}$, we obtain
\begin{equation}\label{stateq}
w_{\rm \Lambda}=-(\frac{1}{3}+\frac{2\sqrt{\Omega_{\rm
\Lambda}}}{3c}\cos y+\frac{\Gamma}{3H}).
\end{equation}
Here as in Ref.\cite{WGA}, we choose the following relation for
decay rate
\begin{equation}\label{decayeq}
\Gamma=3b^2(1+r)H
\end{equation}
with  the coupling constant $b^2$. Using Eq.(\ref{ratio}), the
above decay rate take following form
\begin{equation}\label{decayeq2}
\Gamma=3b^2H\frac{(1+\Omega_{k})}{\Omega_{\Lambda}}
\end{equation}
Substitute this relation into Eq.(\ref{stateq}), one finds the
holographic energy equation of state
\begin{equation} \label{3eq4}
w_{\rm \Lambda}=-\frac{1}{3}-\frac{2\sqrt{\Omega_{\rm
\Lambda}}}{3c}\cos y-\frac{b^2(1+\Omega_{k})}{\Omega_{\rm \Lambda}}.
\end{equation}
If we take $c=1$, then $w_{\Lambda}$ is bounded from below for a
fixed $b^2$ by
\begin{equation} \label{3eq41}
w_{\rm \Lambda}=-\frac{1}{3}(1+2\sqrt{\Omega_{\rm
\Lambda}}+\frac{3b^2(1+\Omega_{k})}{\Omega_{\rm \Lambda}}).
\end{equation}
  According to relation $y\equiv \sqrt{k}R_h/a$, $\cos
y=1$ when $k=0$, in this case $\Omega_{k}=0$ also, therefore in flat
universe, the holographic energy equation of state take following
form
\begin{equation} \label{3eq40}
w_{\rm \Lambda}=-\frac{1}{3}-\frac{2\sqrt{\Omega_{\rm
\Lambda}}}{3c}-\frac{b^2}{\Omega_{\rm \Lambda}}.
\end{equation}
which is exactly the result of \cite{Kim:2005at}. From
Eqs.(\ref{eff}, \ref{decayeq2}, \ref{3eq4}), we have the effective
equation of state as
\begin{equation} \label{3eq401}
w_{\rm \Lambda}^{eff}=-\frac{1}{3}-\frac{2\sqrt{\Omega_{\rm
\Lambda}}}{3c}\cos y.
\end{equation}
If we take $c=1$, and taking $\Omega_{\Lambda}=0.73$ for the present
time, the lower bound of $w_{\rm \Lambda}^{eff}$ is $-0.9$.
Therefore it is impossible to have $w_{\rm \Lambda}^{eff}$ crossing
$-1$. This implies that one can not generate phantom-like equation
of state from an interacting holographic dark energy model in
non-flat universe.
 \section{Conclusions}
In order to solve cosmological problems and because the lack of our
knowledge, for instance to determine what could be the best
candidate for DE to explain the accelerated expansion of universe,
the cosmologists try to approach to best results as precise as they
can by considering all the possibilities they have. It is of
interest to remark that in the literature, the different scenarios
of DE has never been studied via considering special similar
horizon, as in \cite{davies}, in the standard cosmology framworke,
the apparent horizon, $1/H$, determines our universe while in
\cite{gong}, in the Brans-Dicke cosmology framworke, the universe is
enclosed by event horizon, $R_h$. As we discussed in introduction,
for flat universe the convenient horizon looks to be $R_h$ while in
non-flat universe we define $L$ because of the problems that arise
if we consider $R_h$ or $R_p$ (these problems arise if we consider
them as the system's IR cut-off). In present paper, we studied $L$,
as the horizon measured from the sphere of the horizon as system's
IR cut-off. Then, by considering an interaction between holographic
energy density and CDM, we have obtained the equation of state for
the interacting holographic energy density in the non-flat universe.

\end{document}